\documentclass[aps,pra,twocolumn,groupedaddress]{revtex4-2}  
\usepackage{graphicx}
\usepackage{amssymb}
\usepackage{amsmath}
\usepackage{newtxtext,newtxmath,lipsum}

\usepackage[shortlabels]{enumitem}

\begin{document}
\bibliographystyle{revtex}

	\title{Quantum Synchronization in quadratically coupled quantum van der Pol oscillators}

	\author{Nissi Thomas}
	\affiliation{Department of Nonlinear Dynamics, School of Physics, Bharathidasan University \\ 
		Tiruchirappalli - 620024, Tamil Nadu, India.}
	
	\author{M. Senthilvelan}
	\affiliation{Department of Nonlinear Dynamics, School of Physics, Bharathidasan University \\ 
		Tiruchirappalli - 620024, Tamil Nadu, India.}

	\begin{abstract}
		We implement nonlinear anharmonic interaction in the coupled van der Pol oscillators  to investigate the quantum synchronization behaviour of the systems. We study the quantum synchronization in two oscillator models, coupled quantum van der Pol oscillators and anharmonic self-oscillators.  We demonstrate that the considered systems exhibit a high-order synchronization through coupling in both classical and quantum domains. We show that due to the anharmonicity of the nonlinear interaction between the oscillators the system exhibits phonon blockade in the phase locking regime which is a pure nonclassical effect and has not been observed in the classical domain. We also demonstrate that for coupled anharmonic oscillators the system shows a multiple resonance phase locking behaviour due to nonlinear interaction. We point out that the synchronization blockade arises due to strong anticorrelation between the oscillators  which leads to phonon antibunching in the same parametric regime. In the anharmonic oscillator case we illustrate the simultaneous occurrence of bunching and antibunching effects as a consequence of simultaneous negative and positive correlation between the anharmonic oscillators. We examine the aforementioned characteristic features in the frequency entrainment of the oscillators using power spectrum where one can observe normal mode splitting and Mollow triplet in the strong coupling regime. Finally, we propose a possible experimental realization for the considered system in trapped ion and optomechanical settting.
	\end{abstract}

\maketitle
\section{INTRODUCTION}\label{intro}
Synchronization is the adjustment of rhythms of coupled self-sustained systems around a common frequency, which were once independent systems with different frequencies \cite{pikovsky}. It is a ubiquitous phenomenon occurring in different physical, biologial and chemical systems such as neuronal networks, power-grid networks, rhythm circadian in mammmals, electrical circuits, lasers, orbital resonances in planetary systems and so on. Some noted examples with interesting application of synchronization are heart cardiac pacemaker cells, chaotic laser signals and micromechanical oscillators \cite{varela,fell,miller,antonia,arroyo}. Synchronization is a well-understood phenomenon in classical systems and it has been studied in different contexts. For example self-sustained oscillators with external drive, two coupled systems and globally coupled systems with random frequencies such as Kuramoto model \cite{kuramoto,acebron}.
\par Recent developments of quantum systems such as nanomechanical oscillators \cite{aspelmeyer}, superconducting circuits \cite{clarke,wendin}, quantum electrodynamics \cite{blais} and Trapped ions \cite{haffner}, have witnessed significant progress. Since these systems exhibit synchronization properties like limit cycle oscillations, nonlinearity and so on, the idea of synchronization in quantum regime emanated. With these developments synchronization effects in different quantum systems such as atomic ensembles \cite{xu,zhu,cabot}, Josephson circuits \cite{nigg}, stochastic systems \cite{goyu}, Kerr-anharmonic oscillators \cite{bruder}, micromasers \cite{armour}, spin systems \cite{roulet1,roulet2,laskar,tindall} have been studied recently. Synchronization behaviours have also been  investigated in experimental platforms such as optomechanical systems \cite{weiss,amitai1,du,fan,geng,liao,ludwig,qiao}, nanomechanical oscillators \cite{matheny} and superconducting devices \cite{hriscu} in the quantum domain. These works shed light on several quantum aspects of quantum synchronization where quantum effects play a dominant role in the synchronicity of the systems. Several measures of synchronization have also been proposed in these works inorder to analyze the synchronization behaviour in the quantum regime \cite{weiss, mari, jaseem}.
\par Van der pol oscillators are self-sustained systems which are simple and excellent models to study synchronization.  Recent works characterize different synchronization behaviours in the   context of quantum van der Pol socillators. In these works, different quantum synchronization behaviours such as limit cycle \cite{lee}, frequency entrainment \cite{walter14,walter}, amplitude death phenomena \cite{amitai2}, quantized synchronization behaviour \cite{bruder} and enhancement of synchronizaton through squeezing \cite{sonar} have been investigated. An effective quantum model has also been proposed to capture the underdamped phase dynamics which helped to identify a quality factor for the quantum coherence \cite{weiss2}. In dissipative coupled quantum van der pol oscillators the existence of entanglement between the coupled oscillators \cite{lee2}, frequency entrainment \cite{walter} and amplitude death \cite{amitai2} has been investigated. Quantum van der Pol oscillator has its relevance in trapped ion experiment as well. Quantum synchronization in the context of trapped-ions has been investigated in \cite{lee}. Trapped ion is an ideal platform for quantum information processing and quantum computations due to their better coherence time and quantum control. The trapped-ions experience nonlinear coulomb interactions between ion modes. The cross-Kerr nonlinear terms arising in the coulomb interaction can be implemented as nonlinear interaction between the ion modes which shifts the normal mode frequency of the ion motion \cite{ding17}. 
\par In this work, we consider such nonlinear coulomb interaction between two van der Pol oscillators. In nonlinear susceptible materials this kind of interaction is called $\chi^{(2)}$ nonlinearity and are known to exhibit nonclassical effects like  phonon or photon blockade and strong anti-correlation between photon or phonons in optical systems \cite{xu,lee3,zhou,gerace,xie}. Recently, phonon antibunching was investigated in quantum van der Pol oscillator which significantly depended on two phonon loss \cite{li}.  Quantum correlations due to entanglement have been investigated in the coupled cavities with second-harmonic generation \cite{lee3}. Motivated by the above, in this work, we investigate the effect of nonlinear interaction in phase-locking dynamics of two quantum van der Pol oscillators. We also investigate the phase dynamics of anharmonic self-oscillators with this nonlinear interaction. Using perturbation analysis we obtain expressions for steady state and also for synchronization measure, and analyze the system analytically and numerically. Our results show that due to strong correlations between the oscillators the synchronization peaks suppress at resonance with increasing coupling strength. Further, strong anticorrelations between the oscillators also lead to antibunching. We demonstrate that the nonlinear interaction between the anharmonic self-oscillators causes the system to exhibit multiple resonances in the phase locking regime. Further, we show that in the phase locking parametric regime the oscillators are simultaneously correlated and anticorrelated at different resonances and as a result the system exhibits bunching and antibunching effects simultaneously. We show that these synchronization behaviours are purely nonclassical and have not been observed in the classical regime. Finally, we illustrate the above characteristics using power spectrum where we can observe normal mode splitting and Mollow triplet in the strong coupling regime.
\par We organize the presentation as follows. In Sec. \ref{model}, we describe the system with cross-Kerr interactions using the master equation of the van der Pol oscillator. We also describe the coupled self oscillators and also the coupled anharmonic self-oscillators. In Sec. \ref{classical}, we discuss the steady state dynamics of the system in classical regime. In Sec. \ref{quantum}, we illustrate the synchronization dynamics due to the nonlinear coupling in the quantum regime. We analytically obtain the expression for steady states of the master equation as well as the expression for synchronization measure using perturbation theory in coupling strength and demonstrate the quantum effects in the phase-locking dynamics in the considered system. We also discuss the phonon statistics of the system in the phase-locking parametric regime. In Sec. \ref{power_sp}, we discuss these characteristics in frequency entrainment using power mechanical spectrum. Finally, we summarize our results in Sec. \ref{conc}.

\section{Model}\label{model}
\par
\begin{figure}
	\includegraphics[width=1.0\linewidth]{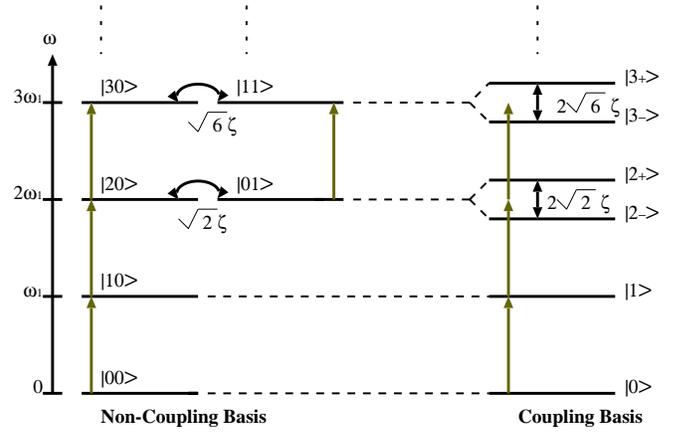}
	\caption{(a) The Schematic Energy level diagram of the nonlinearly coupled quantum van der Pol  oscillator (without Kerr anharmonicity) in the non-coupling(left) and coupling basis (right).}
	\label{Fig1}
\end{figure}

 We consider a nonlinear anharmonic coupling between two quantum van der Pol oscillators. This nonlinear anharmonic interaction is generated by the nonlinear coulomb interaction between two normal modes of motion of two co-trapped ions \cite{ding17}. The master equation governing the time evolution of density matrix $\rho$ of two nonlinearly coupled quantum van der Pol oscillators is described by \cite{lee}  
\begin{equation}
\dot \rho=-i[H_0+H_I,\rho]+\sum_{i=1}^{2}\gamma_1\mathcal{L}[a_i^{\dagger}]\rho+\gamma_2\mathcal{L}[a_i^2]\rho, \label{mas_eq}
\end{equation}
where $a_i(a_i^{\dagger})$ are the annihilation (creation) operators of the $i^{th}$ oscillator. The system Hamiltonian is given by
\begin{equation}
H_0 = \sum_{i=1}^2\omega_ia_i^{\dagger}a_i+K_ia_i^{\dagger 2}a_i^2,\label{sys_ham}
\end{equation} 
where $\omega_i$ is the natural frequency and $K_i$ is the Kerr strength of the $i^{th}$ oscillator. In the absence of Kerr nonlinearity the system has energy spectrum as illustrated in Fig. \ref{Fig1} in the non-coupling basis. The presence of Kerr nonlinearity  in the system Hamiltonian $H_0$ brings the anharmonicity in the energy spectrum and this leads to a spacing of $\omega_i+(m+1)K_i$ between $m^{th}$ and $(m+1)^{th}$ levels of the energy spectrum of the $i^{th}$ oscillator which in turn brings a shift in the energy levels of the Fock states \cite{amitai2}. The Lindbland operator $\mathcal{L}[\hat o]\rho=\hat o\rho o^{\dagger}-\frac{1}{2}\{\hat o^{\dagger}\hat o\rho+\rho\hat o^{\dagger}\hat o\}$ describes the non-unitary dynamics of the system and the parameter $\gamma_1$ in (\ref{mas_eq}) denotes the rate of phonon gain and $\gamma_2$ is the rate of nonlinear phonon loss. As the nonlinear phonon rate ($\gamma_2$) increases the oscillator occupies fewer phonon Fock states. Therefore in the limit $\gamma_2/\gamma_1\to\infty$ the system shows a discrete level structure and this corresponds to the quantum limit where the radius of the limit cycle decreases. On the other hand, if the limit $\gamma_2/\gamma_1\to0$, the radius of the limit cycle increases and the system becomes highly excited.  As a result the system approaches the classical limit.
The nonlinear interaction Hamiltonian $H_I$ is considered in the following form
\begin{equation}
H_I=\zeta(a_1^{\dagger 2} a_2+a_1^2a_2^{\dagger}),\label{int_h}
\end{equation}
with $\zeta$ as the coupling strength. The nonlinear interaction $H_I$ mediates the conversion of the phonon of the first oscillator into two phonons of the second oscillator and vice versa and as a result the eigenenergies of the system changes. We study the dynamics of the system (\ref{mas_eq}) using this nonlinear interaction (\ref{int_h}) in two oscillator models, namely (i) quantum self-oscillators ($K=0$ in Eq. (\ref{sys_ham})) and (ii) anharmonic self-oscillators ($K\neq0$ in Eq. (\ref{sys_ham})). We illustrate the energy spectrum of the Hamiltonian $H_0+H_I$ with $K=0$ in Fig. \ref{Fig1}. In the non-coupling basis $|n_1,n_2 \rangle$ represents the Fock states of the coupled system, where $|n_1\rangle$ and $|n_2\rangle$ corresponds to Fock states of the first and second oscillator respectively. In Fig. \ref{Fig1}, $|00\rangle$ represents the ground state and $|10\rangle$ represents the first excited state. In the absence of the coupling ($\zeta=0$) the bare energy eigenstates $|20\rangle$ and $|01\rangle$ and the energy eigenstates $|30\rangle$ and $|11\rangle$ are degenerate in the second excitation manifold $(n=2)$ and third excitation manifold $(n=3)$ respectively. In the coupling basis $|0\rangle=|00\rangle$ and $|1\rangle=|10\rangle$ represents the ground and first excited states.  In the second excitation manifold ($n=2$) the coupling lifts the degeneracy of the eigenstates $|20\rangle$ and $|01\rangle$ and these eigenstates split into two non-degenerate states $|2_{\pm}\rangle=\frac{1}{\sqrt{2}}(|20\rangle\pm|01\rangle)$ with a separation of $2\sqrt{2}\zeta$. Similarly in the third excitation manifold ($n=3$) the degeneracy between the states $|30\rangle$ and $|11\rangle$ is lifted off, giving rise to two non-degenerate states $|3_{\pm}\rangle=\frac{1}{\sqrt{2}}(|30\rangle\pm|11\rangle)$ with a separation of $2\sqrt{6}\zeta$. In the following, we analyze the phase-locking behaviour in the classical and quantum domains of the coupled systems with $K=0$ and $K\neq0$ in (\ref{sys_ham}) and point out the features that exist only in the quantum regime. 
\section{Synchronization in Classical regime}\label{classical}
\begin{figure*}
	\includegraphics[width=1.0\linewidth]{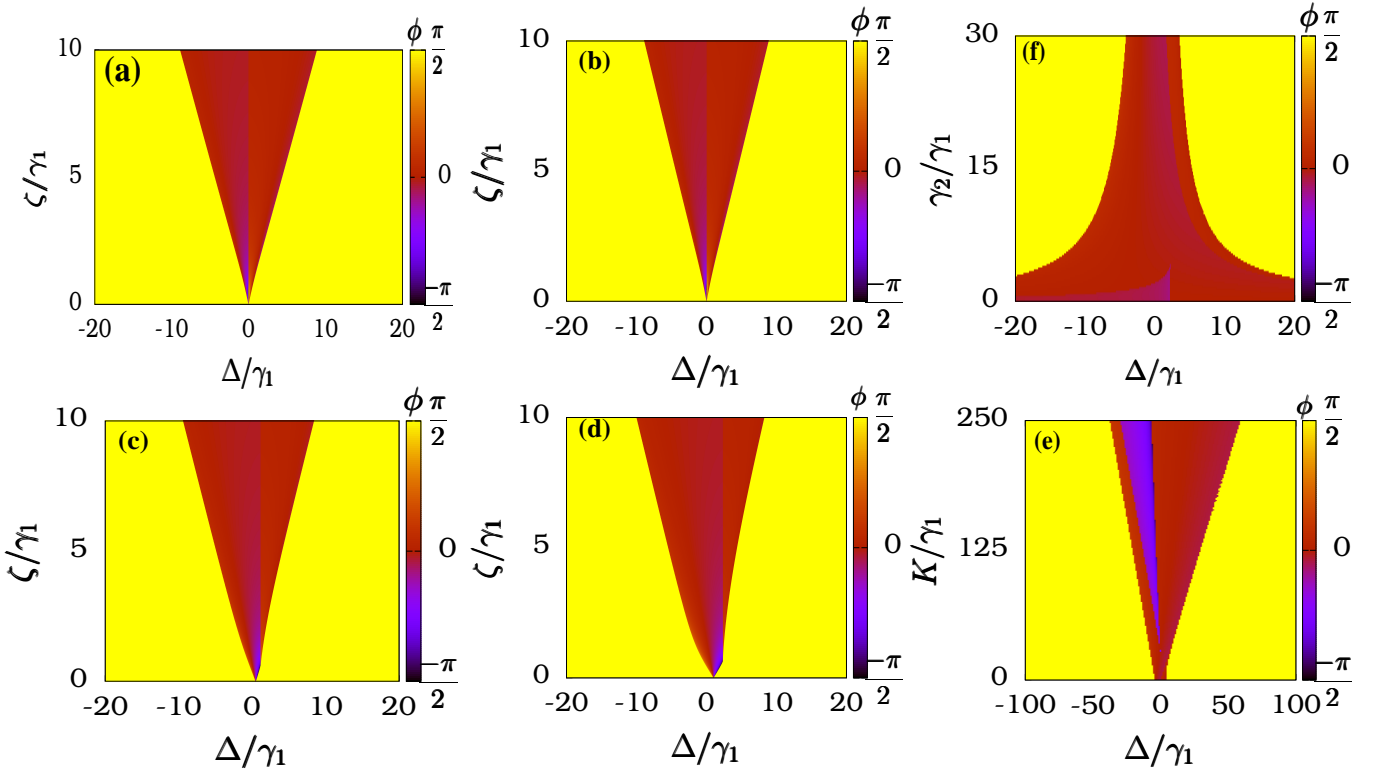}
	\caption{Phase locking behaviour of steady states in classical regime. Synchronization regime plotted as a function $\zeta$ and $\Delta$ for (a) $K=0\gamma_1$, (b) $K=0.1\gamma_1$, (c) $K=5\gamma_1$ and (d) $K=10\gamma_1$ with $\gamma_2/\gamma_1=10$. Fig. \ref{Fig2}(e) illustrates the phase locking behaviour of the steady state as a function of $K$ and $\Delta$ with $\zeta=5$ and $\gamma_2/\gamma_1=10$. Fig. \ref{Fig2}(f) shows the phase locking behaviour as a function of $\gamma_2/\gamma_1$ and $\Delta$ with $K=10$ and $\zeta=5$. In all the figures relative phase $\phi$ is plotted using the color scale.} 
	\label{Fig2}
\end{figure*}
First we consider the system (\ref{mas_eq}) in the limit $\gamma_2/\gamma_1\to0$ where the limit cycle amplitude of the system becomes large. Applying Heisenberg equation of motion followed by mean field approximation, the equations of motion for the limit cycle amplitudes $\alpha_1 = \langle a_1 \rangle$ and $\alpha_2 = \langle a_2 \rangle$ can be formulated as
\begin{eqnarray}
~~~~\dot \alpha_1 &=& (-i\omega_1-2iK|\alpha_1|^2+\frac{\gamma_1}{2}-\gamma_2|\alpha_1|^2)\alpha_1-2i\zeta\alpha_1^*\alpha_2, \nonumber \\
\dot \alpha_2 &=& (-i\omega_2-2iK|\alpha_2|^2+\frac{\gamma_1}{2}-\gamma_2|\alpha_2|^2)\alpha_2-i\zeta\alpha_1^2.\label{albe}
\end{eqnarray}
\par  Using polar coordinates, $\alpha_1=r_1\exp(i\theta_1)$ and $\alpha_2=r_2\exp(i\theta_2)$,  Eq. (\ref{albe}) can be rewritten as phase and amplitude equation in the form
	\begin{eqnarray}
	\dot r_1&=&\big(\frac{\gamma_1}{2}-\gamma_2 r_1^2\big)r_1+2\zeta r_1r_2\sin\phi, \nonumber \\
    \dot r_2&=&\big(\frac{\gamma_1}{2}-\gamma_2 r_2^2\big)r_2-\zeta r_1^2\sin\phi, \nonumber \\   
	\dot \phi&=&-\Delta-2K(r_2^2-2r_1^2)-\zeta\big(\frac{r_1^2-4r_2^2}{r2}\big)\cos\phi\label{amph},
\end{eqnarray}
where $\phi=\theta_2-2\theta_1$ is the phase difference and $\Delta=\omega_2-2\omega_1$ is the frequency detuning. From Eq. (\ref{amph}) we can say that the system (\ref{albe}) is synchronized when the frequency of the second oscillator becomes twice the frequency of first oscillator ($\omega_2=2\omega_1$), and a fixed relative phase relation is established between them. Therefore, in the rotating frame, finding the stable fixed point can determine the synchronized regime of the system. In Fig. \ref{Fig2} the synchronized regime (dark colored region showing Arnold tongue) is illustrated which corresponds to a stable fixed point of Eq. (\ref{albe}) and we have plotted relative phase $\phi$ as the color scale in Fig. \ref{Fig2}.

In the steady state regime the amplitude and phase of the system are given by the expressions
\begin{eqnarray}
r_{1}^{*}&=&\sqrt{z}r_{2}^{*},  \quad r_{2}^{*}=\sqrt{\frac{\gamma_1}{2\gamma_2}\frac{(z+2)}{(z^{2}+2)}}, \nonumber \\ \phi^{*}&=&\tan^{-1}\left[-\frac{\gamma_2r_2^{*2}}{\Delta+2Kr_2^{*2}(1-2z)}\frac{(z-1)(z-4)}{(z+2)}\right],\label{ss1}
\end{eqnarray}
where $r_1^{*}, r_2^{*}$ and $\phi^{*}$ represents the steady state amplitudes and phase of the coupled system (\ref{albe}) and the expression of $z$ can be obtained by solving the following quintic polynomial, that is
\begin{eqnarray}
	\frac{\gamma_1}{2\gamma_2}\zeta^{2}(z+2)(z^{2}+2)(z-4)^{2}-\frac{\gamma_1^{2}}{4}(z-1)^{2}(z-4)^{2}\nonumber\\-\left(2K\frac{\gamma_1}{2\gamma_2}(1-2z)(z+2)+\Delta(z^{2}+2)\right)^{2}=0.\label{fpp}
\end{eqnarray}
\par Using linear stability analysis we find that out of five possible roots of $z$ only two are stable and as we increase the value of $K$ one of these stable stationary states becomes unstable.  In Fig. \ref{Fig2} we illustrate the region where the nonlinearly coupled system (\ref{albe}) is synchronized corresponding to the stationary states given in Eqs. (\ref{ss1}) and (\ref{fpp}) for different values of $K$. Figure \ref{Fig2}(a) shows the synchronization regime (or Arnold tongue) of the coupled system for $K=0$. We have studied the dynamics of the system (\ref{albe}) for $K=0$ in detail very recently \cite{nis}. Here we have shown that the system exhibits high-order synchronization and multistable behaviour which arises due to the presence of rotational symmetry in the system. 
For $\Delta=0$ (and $K=0$) we obtain steady state solutions corresponding to $\cos\phi^{*}=0$ and $r_1^{*}=2r_2^{*}$ from the phase equation given in Eq. (\ref{amph}). For $\cos\phi^{*}=0$, there exists two solutions for $\phi^{*}$, that is $\phi^{*}=\pi/2$ and $\phi^{*}=3\pi/2$. Among these two, only the steady state corresponding to $\phi^{*}=\pi/2$ is stable for $\zeta<\zeta_c$. When we increase the value of the coupling strength, beyond a critical coupling strength ($\zeta>\zeta_c$), the steady state solution corresponding to $\phi^{*}=\pi/2$ loses its stability. One may find that the following two stable solutions for $\phi^{*}$ \cite{nis}, 
\begin{eqnarray}
\phi^{*}&=&\phi_0 \quad \text{and} \quad \phi^{*}=-\phi_0+\pi, \nonumber\\
\phi_0&=&\sin^{-1}\left({\frac{1}{2\zeta}}\sqrt{\frac{\gamma_1\gamma_2}{6}}\right), \label{phi1}
\end{eqnarray} 
corresponding to $r_1^{*}=2r_2^{*}$ arise from the critical point. For both the values of $\phi^{*}$, since the solutions given in Eq. (\ref{phi1}) are stable, the system exhibits a multistable behaviour and as a result the system exhibits clockwise and anticlockwise rotations in the same periodic orbit. Therefore the values of $\phi^{*}$ corresponding to lower values of coupling strength ($\pi/2$)  and higher values of coupling strength (given in Eq. (\ref{phi1})) determines the synchronization regime for $\Delta=0$ which is presented in Fig. \ref{Fig2}(a). For $\Delta\neq0$, the phase deviates from the aforementioned values and attains values corresponding to one of the stable stationary states given in Eq. (\ref{ss1}) (for $K=0$) and the Arnold tongue is obtained as shown in Fig. \ref{Fig2}(a). For $\Delta\neq0$ the system exhibits multistability and oscillators move along the clockwise and anticlockwise directions in different periodic orbits. For $K\neq0$, we demonstrate the synchronization regime of the system (\ref{albe}) in Figs. \ref{Fig2}(b)-\ref{Fig2}(d). As mentioned earlier, the system exists in two stable stationary states and displays clockwise and anticlockwise motions in two periodic orbits. As we increase the value of $K$, one of the stable stationary states becomes unstable. From Eq. (\ref{ss1}) we can infer that for lower values of $K$, the tip of the Arnold tongue coincides with $\Delta=0$ as shown in Fig. \ref{Fig2}(b). For higher values of $K$ there is a shift in the tip of the Arnold tongue from $\Delta=0$ as shown in Figs. \ref{Fig2}(c) and \ref{Fig2}(d). The phase locking also increases with increasing Kerr strength for a range of $\Delta$ values as shown in Fig. \ref{Fig2}(e). The damping parameter $\gamma_2/\gamma_1$ rescales the synchronization regime as presented in Fig. \ref{Fig2}(f) which enters into the steady state through $r_1^{*}$ and $r_2^{*}$. 
 \section{Synchronization in Quantum Regime}\label{quantum}
	 Now we explore the dynamical features of the system (\ref{mas_eq}) that comes out due to the presence of nonlinear coupling in the quantum  limit ($\gamma_2/\gamma_1\to\infty$). The phase locking behaviour in the classical regime discussed in Sec. III is also maintained in the quantum regime with stronger phase locking features \cite{lee}. In this section, we demonstrate certain quantum features that exist in the system due to the presence of coupling.  To begin, we analytically obtain the steady state approximation of the master equation (\ref{mas_eq}) using perturbation theory and derive the synchronization measure in order to gain some analytical understanding about the synchronization behaviour in the quantum regime.
	 \begin{figure*}[!ht]
	 	\includegraphics[width=0.48\linewidth]{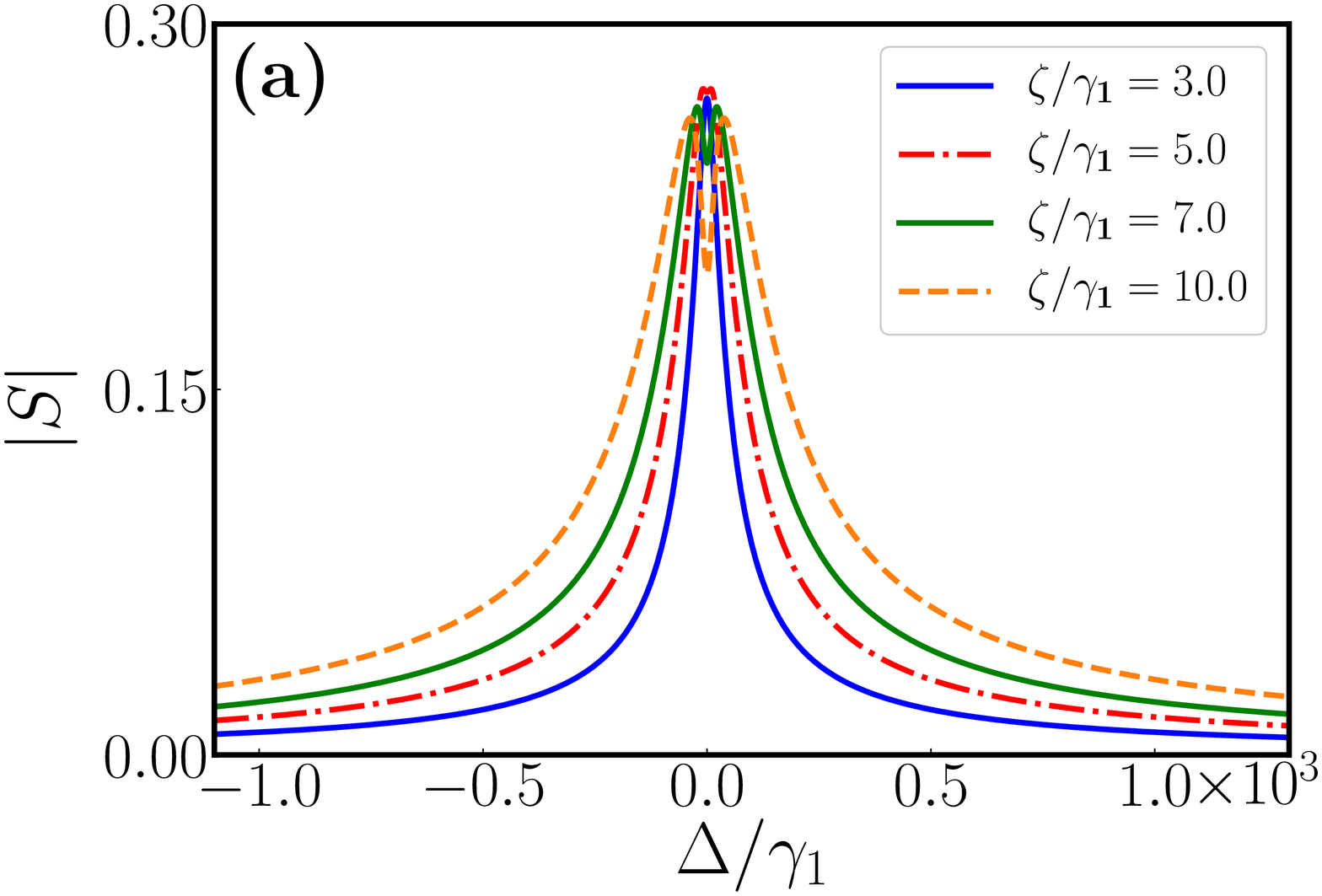}
	 	\includegraphics[width=0.48\linewidth]{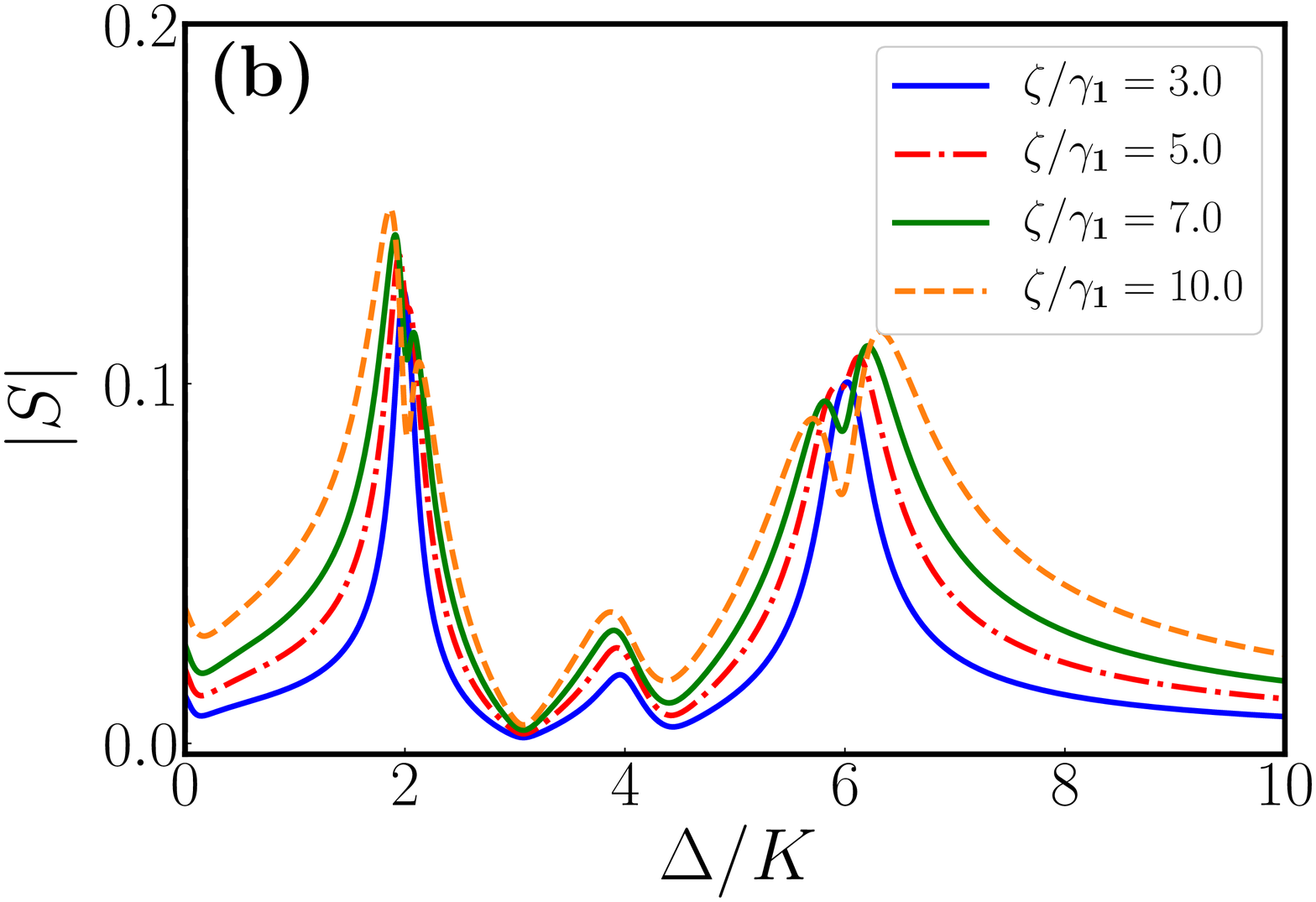}
	 	\caption{Synchronization measure $|S|$ for coupled system (\ref{mas_eq}) plotted as a function of frequency detuning $\Delta$  with different coupling strengths for (a) $K=0$ and (b) $K=250\gamma_1$ with  damping rate $\gamma_2/\gamma_1=10$.}
	 	\label{SMf}
	 \end{figure*}
	 \subsection{Perturbation Analysis}
	 The steady state density matrix of the uncoupled oscillators can be factorized as $\rho^{(0)}=\rho_1^{(0)}\otimes\rho_2^{(0)}$, where $\rho_i^{(0)}$ has a diagonal form and is given by the expression
	\begin{equation}
	\rho_i^{(0)}=\frac{[(\gamma_1/\gamma_2)^{n}\Theta(1+n,\gamma_1/\gamma_2+n,\gamma_1/\gamma_2)]}{[(\gamma_1/\gamma_2)_n\Theta(1,\gamma_1/\gamma_2,2\gamma_1/\gamma_2)]},~~i=1,2\label{ssq}
	\end{equation} 
	where $(\gamma_1/\gamma_2)_n$ denotes the Pochammer symbol and $\Theta$ is the Kummer's confluent hypergeometric function \cite{dodo}.  When the oscillators are uncoupled the quantum van der Pol oscillators are said to exhibit limit cycle oscillations both in the absence  ($K=0$)  and presence ($K\neq0$) of Kerr nonlinearity \cite{amitai2}. This is clear from Eq. (\ref{ssq}) since $\rho_i^{(0)}$ only depends on the parameter $\gamma_1/\gamma_2$. Therefore, in the limit $\gamma_2/\gamma_1\to\infty$, $\rho_i^{(0)}$ can be approximated as $\rho_i^{(0)}\to\frac{2}{3}|0\rangle\langle0|+\frac{1}{3}|1\rangle\langle1|+\mathcal{O}(1/\gamma_2)$.
	 Now we apply perturbation theory in order to obtain the steady state operator of the coupled system. In the weak coupling limit the steady state operator can be expanded as a power series expansion in coupling strength $\zeta$ in the following form \cite{bruder}, that is
	 \begin{equation}
	 \rho=\rho^{(0)}+\zeta\rho^{(1)}+\hdots,\label{pow_ser}
	 \end{equation}
	 where $\rho^{(0)}$ is given in Eq. (\ref{ssq}) and $\rho^{(1)}$ is the first order correction to the density operator. To obtain $\rho^{(1)}$ we decompose the master equation (\ref{mas_eq}) into the perturbation operator $L_I\rho=-i[H_{I},\rho]$ and the unperturbed Lindblandian $L_0\rho=\mathcal{L}\rho-i[H_{0},\rho]$. Therefore the first order correction to the steady state density operator can be defined as $\rho^{(1)}=-L_0^{-1}L_I\rho$, where $L_0^{-1}$ is the Moore-Penrose pseudoinverse of the unperturbed Liouvillian $L_0$ \cite{petru}. The inverse of the superoperator $L_0$ can be found by inverting the diagonal elements in the off-diagonal subspace such that $L_0^{-1}|n+2,m-1\rangle\langle n,m|=\lambda^{-1}|n+2,m-1\rangle\langle n,m|$ with
	 \begin{equation}
	 \lambda=i(\Delta-2K(2n-m+2))-\Gamma,\label{eigen}
	 \end{equation}
	 where $\Delta=\omega_2-2\omega_1$ is the frequency detuning and $\Gamma=\frac{\gamma_1}{2}(2(n+m)+5)+\gamma_2((n+2)^2+(m-1)^2-2(n+3))$. Hence the first order correction to the density matrix  can be obtained in the form
	 {\small \begin{equation}
	 \rho_{n+2,m-1;n,m}^{(1)}=\sum_{n=0,m=1}^{\infty}\frac{i\zeta\sqrt{(m)(n+1)(n+2)}(\rho_{n,m}^{(0)}-\rho_{n+2,m-1}^{(0)})}{\lambda}.\label{fcd}
	 \end{equation}}
\par Since the degeneracy of the eigenstates of the system is lifted up due to the coupling as shown in Fig. \ref{Fig1}, it leads to a shift in the eigenstate that can block the transition of phonons for finite detuning in the absence of Kerr nonlinearity ($K=0$) and multiple resonances in the presence of Kerr nonlinearity ($K\neq0$) for the first order response to the coupling strength. In the following sub-section, we will discuss the significance of these two effects in the phase locking behaviour of the system (\ref{mas_eq}).
\subsection{Phase-locking Measure}
 \par Classically, we identified  the relative phase between the oscillators as $\phi=\theta_2-2\theta_1$ (Eq. (\ref{amph})). To quantify the synchronization in the quantum domain, we define the correlator $\langle a_1^{\dagger}a_1^{\dagger}a_2\rangle$ as the measure of relative phase between the coupled system, in which $\langle a_j\rangle=|a_j|e^{-i\theta_j}$ $(j=1,2)$ determines the phase of uncoupled oscillators. Thus, we define the absolute value of phase synchronization measure in the form \cite{talitha} 
 \begin{eqnarray}
 S&=&|S|e^{-i\phi}=\frac{\langle a_1^{\dagger}a_1^{\dagger}a_2\rangle}{\sqrt{\langle a_1^{\dagger}a_1\rangle\langle a_2^{\dagger}a_2}\rangle}\nonumber\\&=&\sum_{nm}\frac{\sqrt{m(n+1)(n+2)}\rho_{n+2,m-1}}{\sqrt{\langle a_1^{\dagger}a_1\rangle\langle a_2^{\dagger}a_2\rangle}}, \label{SM}
 \end{eqnarray} 
  where $\phi=\theta_2-2\theta_1$ is the relative phase difference between the quantum oscillators.  By substituting the expression (\ref{fcd}) into (\ref{SM}) we can obtain the synchronization measure for the first-order correction of density matrix $\rho^{(1)}$ in the form
 \begin{eqnarray}
 S(\rho^{(1)})=\sum_{nm}\left(\rho_{nm}^{(0)}-\rho_{n+2 m-1}^{(0)}\right)\frac{i\zeta(m(n+1)(n+2))}{\lambda\sqrt{\langle a_1^{\dagger}a_1\rangle\langle a_2^{\dagger}a_2\rangle}},\label{eSM}
 \end{eqnarray}
 where $\lambda$ is given in Eq. (\ref{eigen}).
  \par For $K=0$, $S(\rho^{(1)})$ given in Eq. (\ref{eSM}) turns out to be the sum over terms in Eq. (\ref{eigen}) at $\Delta=0$ and around $\Delta=\pm2\sqrt{2}\zeta$ ($\pm2\sqrt{6}\zeta$ for higher phonon transition) for $\zeta\ll\gamma_2/\gamma_1$ and $\zeta<\gamma_2/\gamma_1$ respectively of width $\Gamma$. For $K\neq0$, the expression for synchronization measure is a coherent sum of resonances at $\Delta=2K(2n-m+2)$ and width $\Gamma$. In the limit $\gamma_2/\gamma_1\to\infty$, the resonances are more resolved for $K\gg\Gamma$ but as the limit $\gamma_2/\gamma_1\to0$ the resonances are no longer resolved as we can see from Fig. \ref{Fig2}(e). 
 \begin{figure*}[!ht]
 	\centering
 	\includegraphics[width=1.0\linewidth]{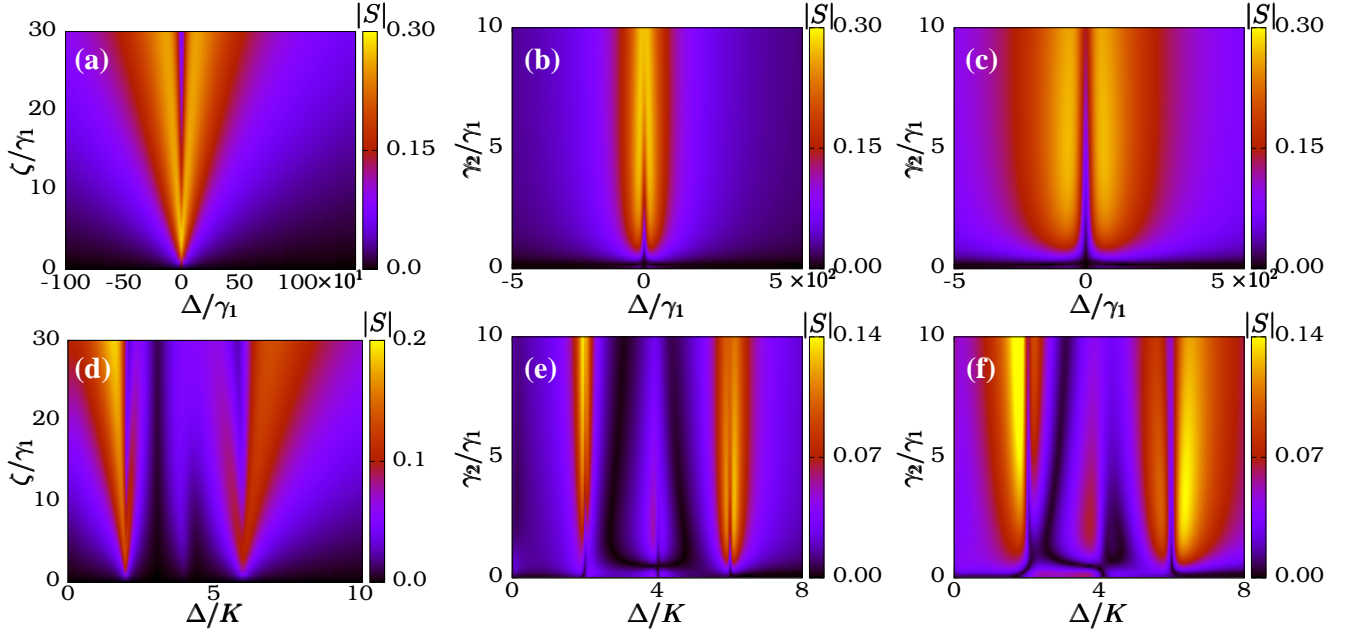}
 	\caption{Different behaviours of the phase locking measure $S$ for the steady state. Synchronization measure $|S|$ plotted as a function of $\zeta$ and $\Delta$ in Figs. \ref{Fig3}(a) with $K=0$ and \ref{Fig3}(d) with $K=250\gamma_1$ for  $\gamma_2/\gamma_1=10$. The same measure $|S|$ is plotted as a function of $\gamma_2$ and $\Delta$ with $K=0$ in Figs. \ref{Fig3}(b)  for $\zeta=5\gamma_1$, \ref{Fig3}(c) for $\zeta=15\gamma_1$ and with $K=250\gamma_1$ for \ref{Fig3}(e) $\zeta=5\gamma_1$, \ref{Fig3}(f) $\zeta=15\gamma_1$   .}
 	\label{Fig3}
 \end{figure*}
\par In Fig. \ref{SMf}, we plot the absolute value of synchronization measure $S$ as a function of $\Delta$ for different coupling strengths using the steady state solver of QuTIP \cite{johan1,johan2}. In Fig. \ref{SMf}(a) we present the synchronization measure for $K=0$. We can observe that for very weak coupling strength $\zeta<<\gamma_2/\gamma_1$, the system exhibits resonance peaks at $\Delta=0$ and as the coupling strength is increased the phase synchronization is suppressed at the resonance $\Delta=0$ with some finite value such that $|S|$ has a local minima and we observe a split in the synchronization peak around $\Delta=0$. As we increase the coupling strength the suppression at the resonance also increases. This can be understood from the energy level diagram given in Fig. \ref{Fig1}. For very low values of $\zeta$ we observe $|0\rangle$ to $|1\rangle$ transition with resonant phonon absorption at $\omega_2=2\omega_1$ and as a result we obtain single phase synchronzation peaks at $\Delta=0$. As we increase the coupling strength,  transition from $|1\rangle$ to $|2_+\rangle$ gets blocked for detuning $\sqrt{2}\zeta$, hence the synchronization is suppressed at $\Delta=0$ and we obtain synchronization peaks at $\Delta=\pm2\sqrt{2}$. Any further increase in the coupling strength $\zeta$ also blocks the $|2_{\pm}\rangle$ to $|3_{\pm}\rangle$ transition for detuning ($\sqrt{6}-\sqrt{2}$)$\zeta$. This leads to an increase in the suppression of phase synchronization and a split in the synchronization peaks as demonstrated in Fig. \ref{SMf}(a). In Fig. \ref{SMf}(b), we illustrate the phase synchronization measure as a function of $\Delta$ for $K\neq0$.  Here we observe that the system exhibits multiple resonances which is clear from Eq. (\ref{eSM}). The reason for the occurrence of multiple resonances in the presence of Kerr anharmonicity in the energy spectrum of the quantum van der Pol oscillators can be explained by considering the multiple phonon transitions due to the nonlinear interaction term present in Eq. (\ref{int_h})  where the creation (annihilation) of two phonons is accompanied by annihilation (creation) of a phonon.  This allows a resonant interaction between the states $|n+2,m-1\rangle$ and $|n,m\rangle$ where $|n\rangle$ and $|m\rangle$ are the Fock states of the first and second oscillator. Thus it is required that the energy eigenvalues satisfies the condition $E(|n,m\rangle)=E(|n+2,m-1\rangle)$ which corresponds to $H_0|n,m\rangle=H_0|n+2,m-1\rangle$ such that we obtain a resonance condition of the following form
 \begin{equation}
 \Delta+2K(2n-m)\pm 4K=0. \label{res_con}
 \end{equation}

 In \cite{bruder} the authors have obtained a resonance condition which has confirmed that coupled identical oscillators (with equal amplitudes) with Kerr anharmonicity in the energy spectrum can show synchronization blockade and the synchronization can be enhanced by making the oscillators more heterogenous. For the interaction given in Eq. (\ref{int_h}), classically, we have observed that the system exhibits high-order synchronization and for $K=0$ the synchronization is maximum for the resonance conditions $\omega_2=2\omega_1$ and $r_1^{*}=2r_2^{*}$ \cite{nis}. For $K\neq0$ the maximal synchronization occurs for the resonance condition $\omega_2=2\omega_1$ and amplitude ratio $r_1^{*}:r_2^{*}=\sqrt{z}:1$ (where $z$ can take range of parametric values) as given in Eq. (\ref{ss1}). Therefore it is clear from the resonance condition (\ref{res_con}) that the system is heterogenous and exhibits multiple resonances at $\Delta=2K(2n-m+2)$.
  \begin{figure}[!ht]
 	\centering
 	\includegraphics[width=1.01\linewidth]{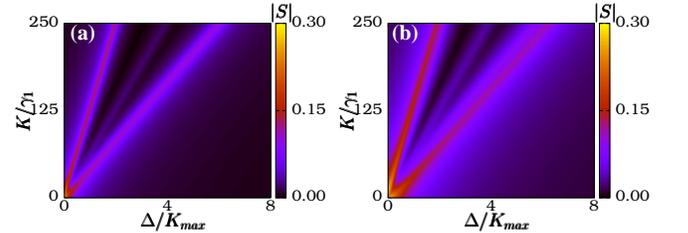}
 	\caption{ Absolute value of synchronization measure $S$  plotted as a function of frequency detuning $\Delta$ and  Kerr parameter $K$ for the damping rate $\gamma_2/\gamma_1=10$ with coupling strength (a) $\zeta=5.0\gamma_1$ and (b) $\zeta=7.0\gamma_1$ and $K_{max}=250\gamma_1$.  }
 	\label{Fig4}
 \end{figure}
\par In Fig. \ref{Fig3} and Fig. \ref{Fig4}, we present an overall picture of the synchronization measure $|S|$ of the steady state (\ref{pow_ser}). Figures \ref{Fig3}(a) and \ref{Fig3}(d), illustrate the synchronization regimes for $K=0$ and $K\neq0$ respectively. In these two figures, we plot the phase-locking measure $|S|$ as a function of $\Delta$ and $\zeta$. Figure \ref{Fig3}(a) reveals that for very lower values of $\zeta$, there is no blockade, and when we increase the coupling strength we can visualize a split in the synchronization tongue at $\Delta=0$. In Fig. \ref{Fig3}(d) we observe the synchronization tongues for $K\neq0$ and upon increasing the coupling strength we observe a split in the synchronization tongues at $\Delta=2K$ and $\Delta=6K$. We will explain the blockade in more detail in the following sub-section. Classically, we observed from Fig. \ref{Fig2} that there is no blockade in the synchronization tongue at $\Delta=0$ for $K=0$ and no multiple resonances are present for $K\neq0$. Figures \ref{Fig3}(b) and \ref{Fig3}(c) illustrate how the blockade increases for lower values of the damping parameter $\gamma_2/\gamma_1$ at the resonance $\Delta=0$ and $K=0$, as the Fock levels become more populated. For $K\neq0$, we can also observe the blockade for different $\Delta$ values for decreasing $\gamma_2/\gamma_1$ in Figs. \ref{Fig3}(e) and \ref{Fig3}(f). The resonances are more resolved with the increasing values of $K$ for different coupling strengths which can be seen from Figs. \ref{Fig4}(a) and \ref{Fig4}(b). Classically, we observed broadening of the resonance (Fig. \ref{Fig2}(e)) for increasing value of $K$. 
\subsection{Mutual Correlation and antibunching}
In the previous sub-section we observed a blockade due to the presence of anharmonicity in the energy spectrum due to nonlinear interaction between two quantum van der Pol oscillators which is a crucial feature to realize phonon blockade and antibunching in the quantum oscillators. Strong phonon (or photons) correlation between the oscillators causes the system to exhibit limit cycle oscillations such that the system synchronizes with each other leading to bunching and antibunching \cite{lee3}.
\begin{figure}[!ht]
	\includegraphics[width=1.0\linewidth]{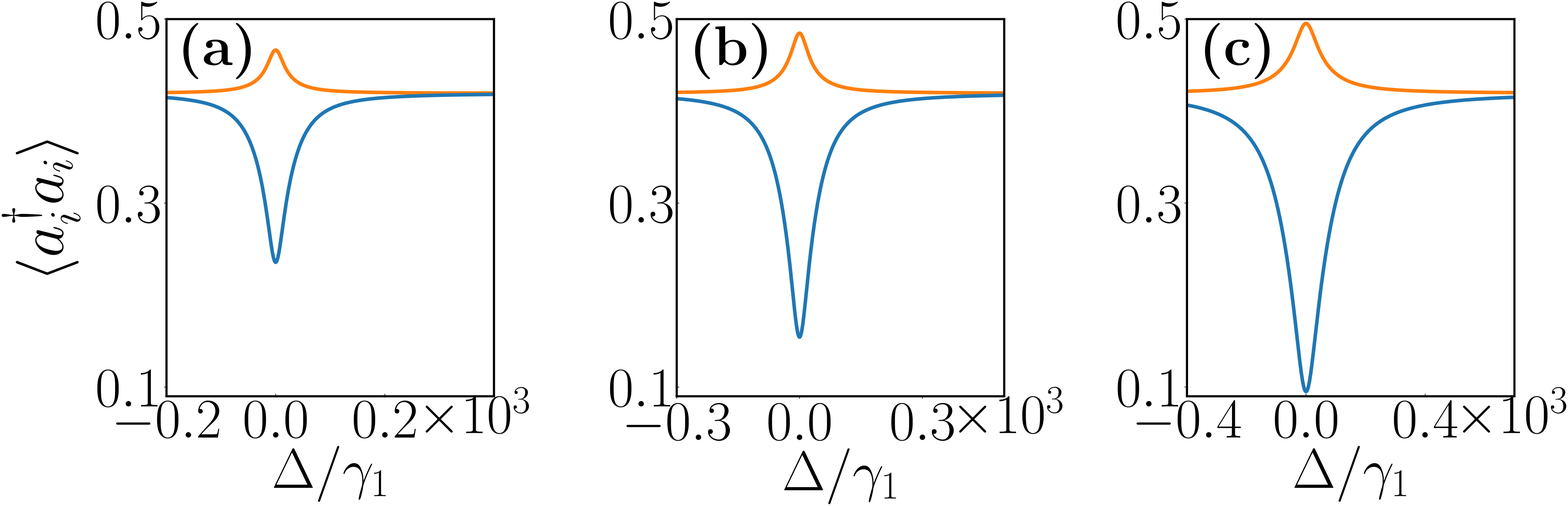}
	\includegraphics[width=1.0\linewidth]{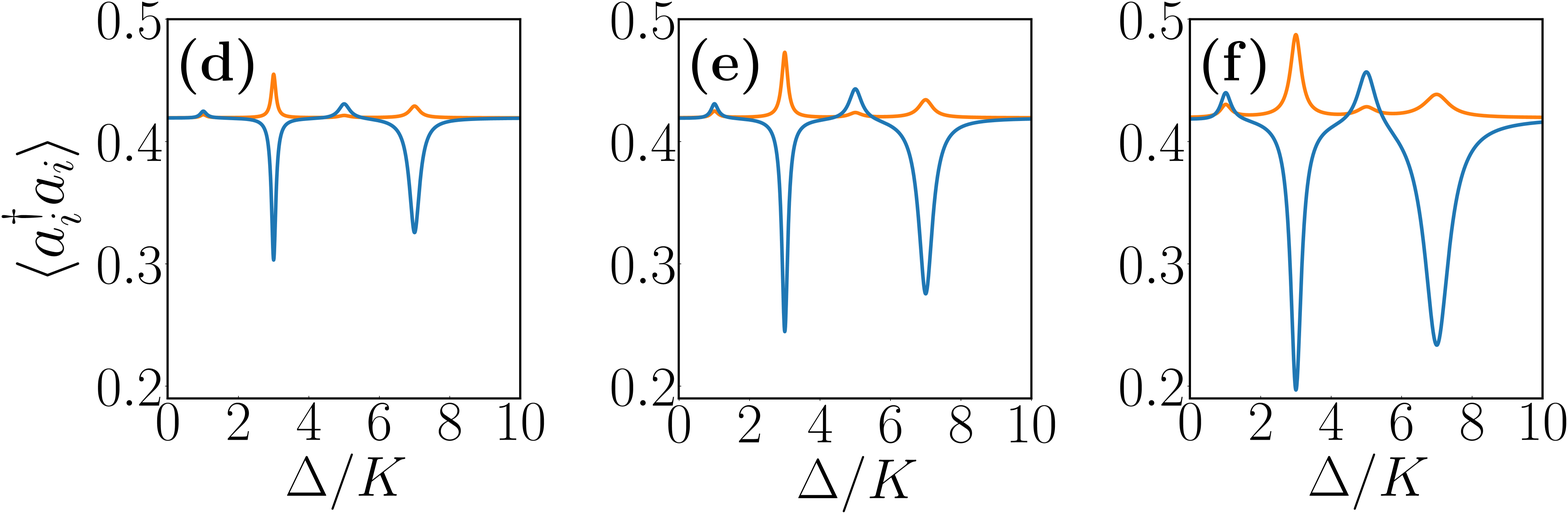}
	\caption{ The phonon number of individual quantum van der Pol oscillators. Here, orange curve represents first oscillator phonon number and blue curve represents second oscillator phonon number. Figures \ref{occup}(a)-\ref{occup}(c) are plotted for $K=0$  with \ref{occup}(a) $\zeta=3\gamma_1$, \ref{occup}(b) $\zeta=5\gamma_1$ and \ref{occup}(c) $\zeta=10\gamma_1$.  Figures \ref{occup}(d)-\ref{occup}(f) are plotted for $K=250\gamma_1$ with \ref{occup}(d) $\zeta=3\gamma_1$, \ref{occup}(e) $\zeta=5\gamma_1$ and \ref{occup}(f) $\zeta=10\gamma_1$. In all these figures we consider $\gamma_2=10\gamma_1$.}
	\label{occup}
\end{figure}  
The phonon correlation between two quantum oscillator modes can be calculated using the second order correlation function \cite{lee3}
\begin{equation}
g_2(a_1,a_2)=\frac{\langle a_1^{\dagger}a_1a_2^{\dagger}a_2\rangle}{\langle a_1^{\dagger}a_1\rangle\langle a_2^{\dagger}a_2\rangle},\label{sec_cor}
\end{equation}
where $g_2$ is the steady state second-order correlation function. The two oscillator modes are positively correlated when $g_2>1$, leading to simultaneous emission of phonons known as bunching. When $g_2<1$ the oscillator modes are negatively correlated and simultaneous emission of phonons are blocked known as antibunching . When $g_2=1$, there are no correlations between the two oscillator modes. Figure \ref{occup} demonstrates the phonon number $\langle a_i^{\dagger}a_i\rangle$ of the first (orange curve) and second (blue curve) oscillator. Figures \ref{occup}(a)-\ref{occup}(c) show the phonon correlation between the first and second oscillator for different values of coupling strength $\zeta$ for $K=0$. We can see that the oscillators are anti-correlated with each other and as the coupling strength is increased the oscillators become more negatively correlated.  
\begin{figure}[!ht]
	\includegraphics[width=0.8\linewidth]{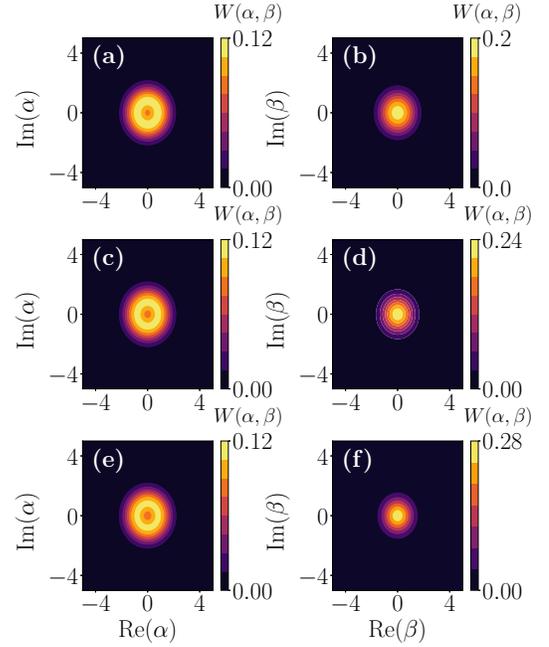}
	\caption{Wigner distribution function for the steady state of the individual oscillators for $K=0$. Figs. \ref{Wigner1}(a), \ref{Wigner1}(c) and \ref{Wigner1}(e) represent the limit cycle of the first oscillator and Figs. \ref{Wigner1}(b), \ref{Wigner1}(d) and \ref{Wigner1}(f) represent the limit cycle of the second oscillator. In  Figs. \ref{Wigner1}(a)-\ref{Wigner1}(b) $\zeta=3\gamma_1$, \ref{Wigner1}(c)-\ref{Wigner1}(d) $\zeta=5\gamma_1$ and \ref{Wigner1}(e) -\ref{Wigner1}(f) $\zeta=10\gamma_1$. In all these figures we consider $\gamma_2=10\gamma_1$ and $\Delta=0$.}
	\label{Wigner1}
\end{figure}
\begin{figure*}[!ht]
	\includegraphics[width=1.0\linewidth]{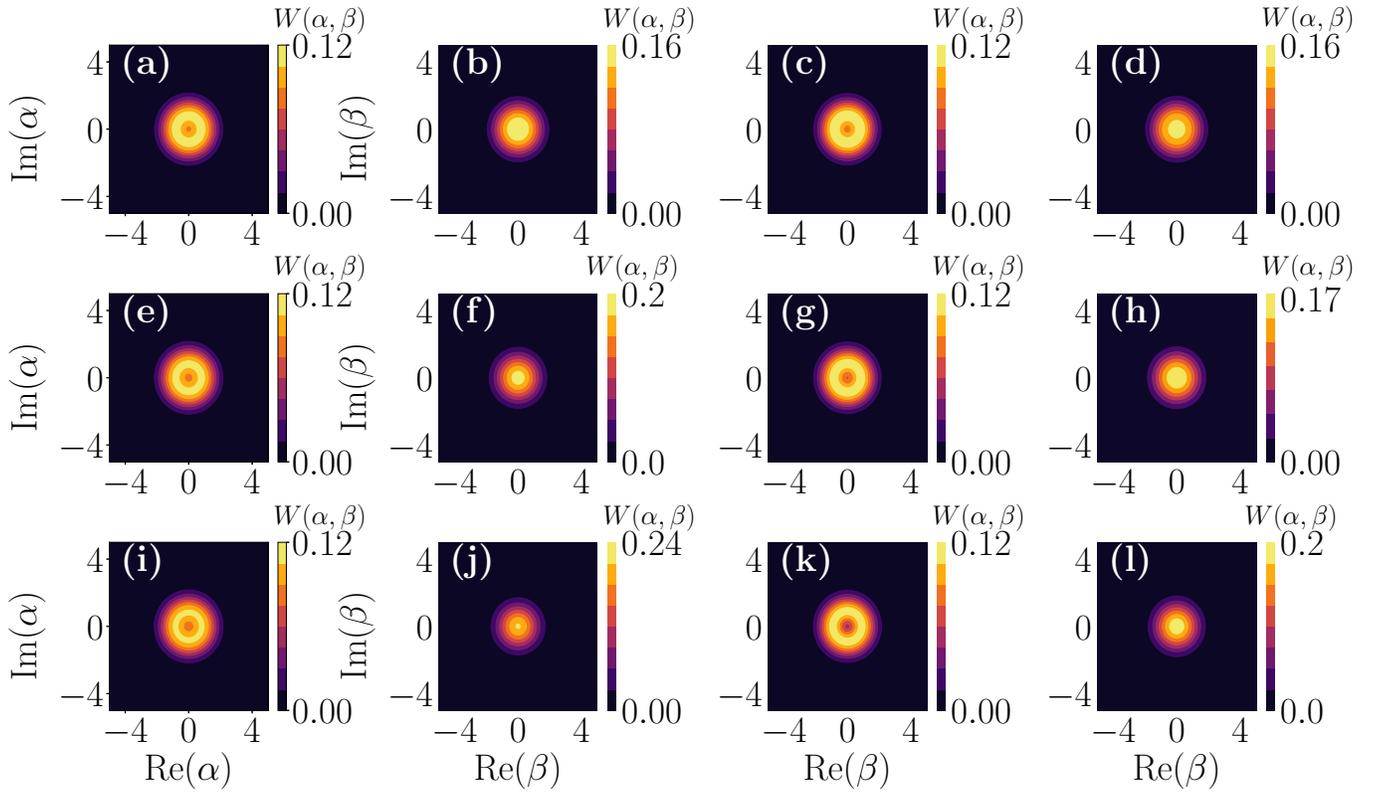}
	\caption{ Wigner distribution function for the steady state of the individual oscillators for $K=250\gamma_1$ and $\gamma_2=10\gamma_1$. Figs. \ref{Wigner1}(a), \ref{Wigner1}(e) and \ref{Wigner1}(i) represent the limit cycle of the first oscillator. The limit cycle of the second oscillator is presented in Figs. \ref{Wigner1}(b) - \ref{Wigner1}(d) for $\Delta=2K$, \ref{Wigner1}(f) - \ref{Wigner1}(h) for $\Delta=4K$, and \ref{Wigner1}(j) - \ref{Wigner1}(l) for $\Delta=6K$. In Figs. \ref{Wigner1}(a) - \ref{Wigner1}(d) $\zeta=3\gamma_1$, \ref{Wigner1}(e)-\ref{Wigner1}(h) $\zeta=5\gamma_1$ and \ref{Wigner1}(i) - \ref{Wigner1}(l) $\zeta=10\gamma_1$. }
	\label{Wigner2}
\end{figure*}
The Wigner dynamics of the first and second oscillator for $\Delta=0$ is plotted for differrent coupling strengths in Fig. \ref{Wigner1}. From  Figs. \ref{occup}(a)-\ref{occup}(c) we can observe that the phonon number increases very slightly with increase in coupling strength at $\Delta=0$ which can be observed in the Wigner function representation in Figs. \ref{Wigner1}(a), \ref{Wigner1}(c) and \ref{Wigner1}(e). We can see that there is no change in the limit cycle of the first oscillator for different $\zeta$ values. In Figs. \ref{occup}(a)-\ref{occup}(c) a negative increase in the phonon number $\langle a_2^{\dagger}a_2\rangle$ is observed which is confirmed from the shrink in the limit cycle as presented in the Wigner distribution function of the second oscillator in Figs. \ref{Wigner1}(b), \ref{Wigner1}(d) and \ref{Wigner1}(f). For $K\neq0$, the phonon number of the first and second oscillators are plotted in Figs. \ref{occup}(d)-\ref{occup}(f). In these figures we can observe the phonon number peaks of the two oscillators at different resonances which are simultaneously correlated and anti-correlated, and this correlation and anti-correlation increases with increasing coupling strength. In Figs. \ref{occup}(d)-\ref{occup}(f) we can also observe that the phonon numbers of the first and second oscillators are negatively correlated at $\Delta=2K$ and $\Delta=6K$ and positively correlated at $\Delta=4K$. 
The Wigner function distribution of the first and second oscillator for different resonance conditions (horizontally) and different coupling strengths (vertically) are plotted in Fig. \ref{Wigner2}. The limit cycle of the first oscillator remains same for all values of $\zeta$ and $\Delta$ which can be seen from Figs. \ref{Wigner2}(a), \ref{Wigner2}(e) and \ref{Wigner2}(i). The limit cycle of the second oscillator is illustrated in Figs. \ref{Wigner2}(b)-\ref{Wigner2}(d) for $\zeta=3\gamma_1$, Figs. \ref{Wigner2}(f)-\ref{Wigner2}(h) for $\zeta=5\gamma_1$ and Figs. \ref{Wigner2}(j)-\ref{Wigner2}(l) for $\zeta=10\gamma_1$. We can observe that for anti-correlated phonons the  limit cycle of the second oscillator shrinks as presented in Figs. \ref{Wigner2}(b), \ref{Wigner2}(f) and \ref{Wigner2}(j) for $\Delta=2K$ and Figs. \ref{Wigner2}(d), \ref{Wigner2}(h) and \ref{Wigner2}(l) for $\Delta=6K$. For positively correlated phonons the limit cycle of the first and second oscillators remains the same as shown in Figs. \ref{Wigner2}(c), \ref{Wigner2}(g) and \ref{Wigner2}(k) for $\Delta=4K$. 

With the results now at hand, we analyze the mutual correlation between the quantum van der Pol oscillators through the second order correlation function given in Eq. (\ref{sec_cor}). 
 \begin{figure}[!ht]
 	\includegraphics[width=0.48\linewidth]{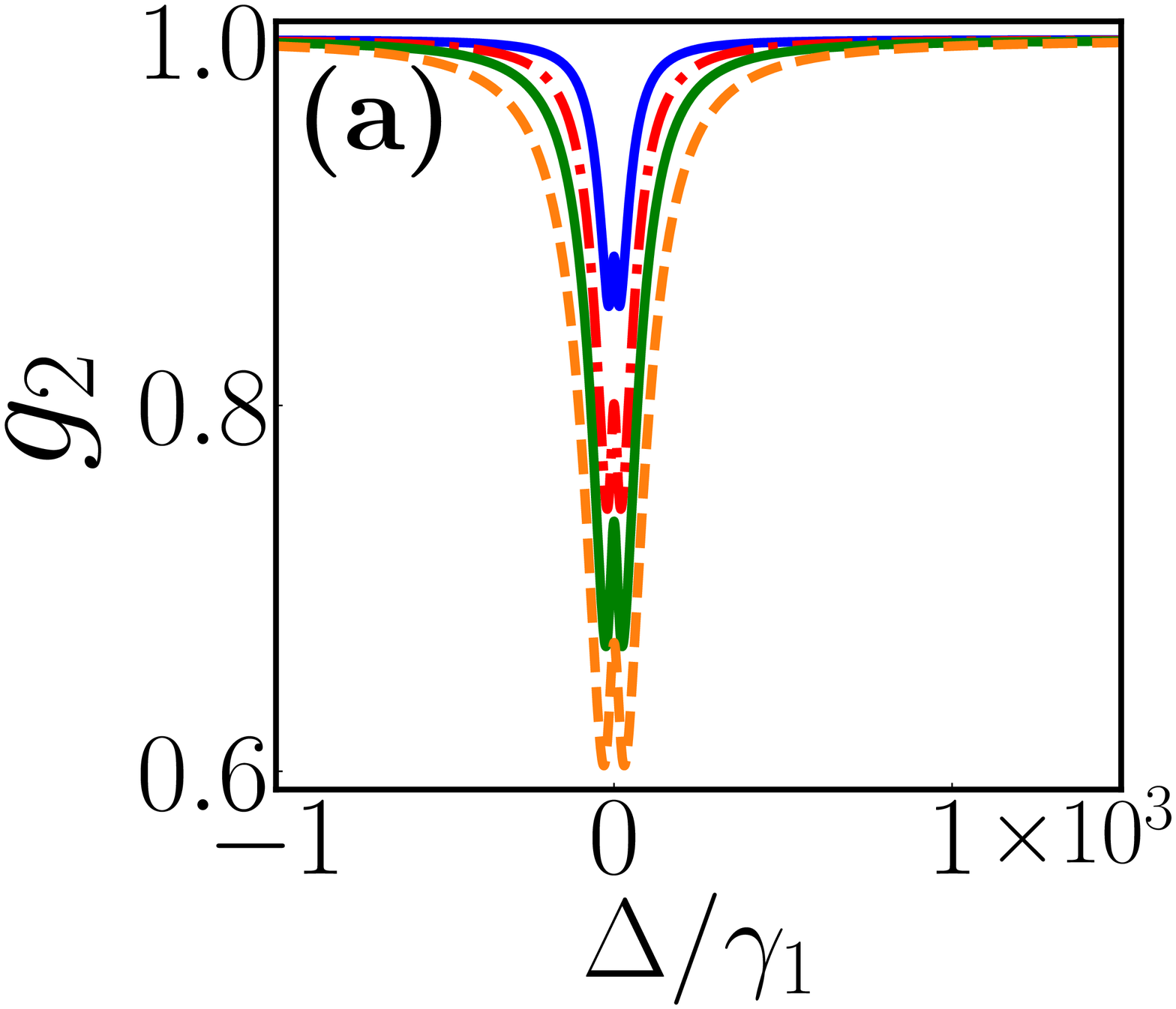}
 	\includegraphics[width=0.48\linewidth]{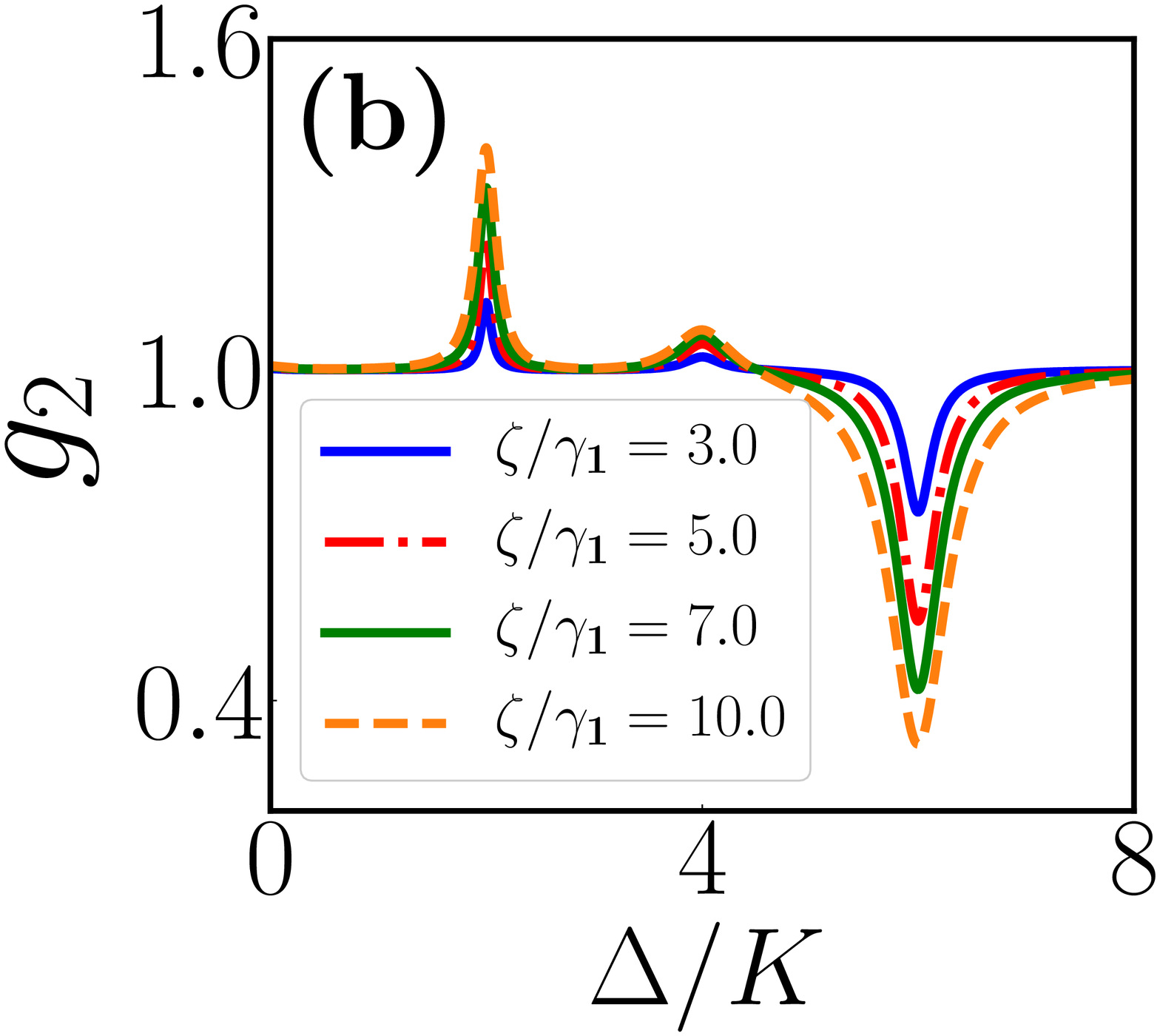}
 	\includegraphics[width=0.48\linewidth]{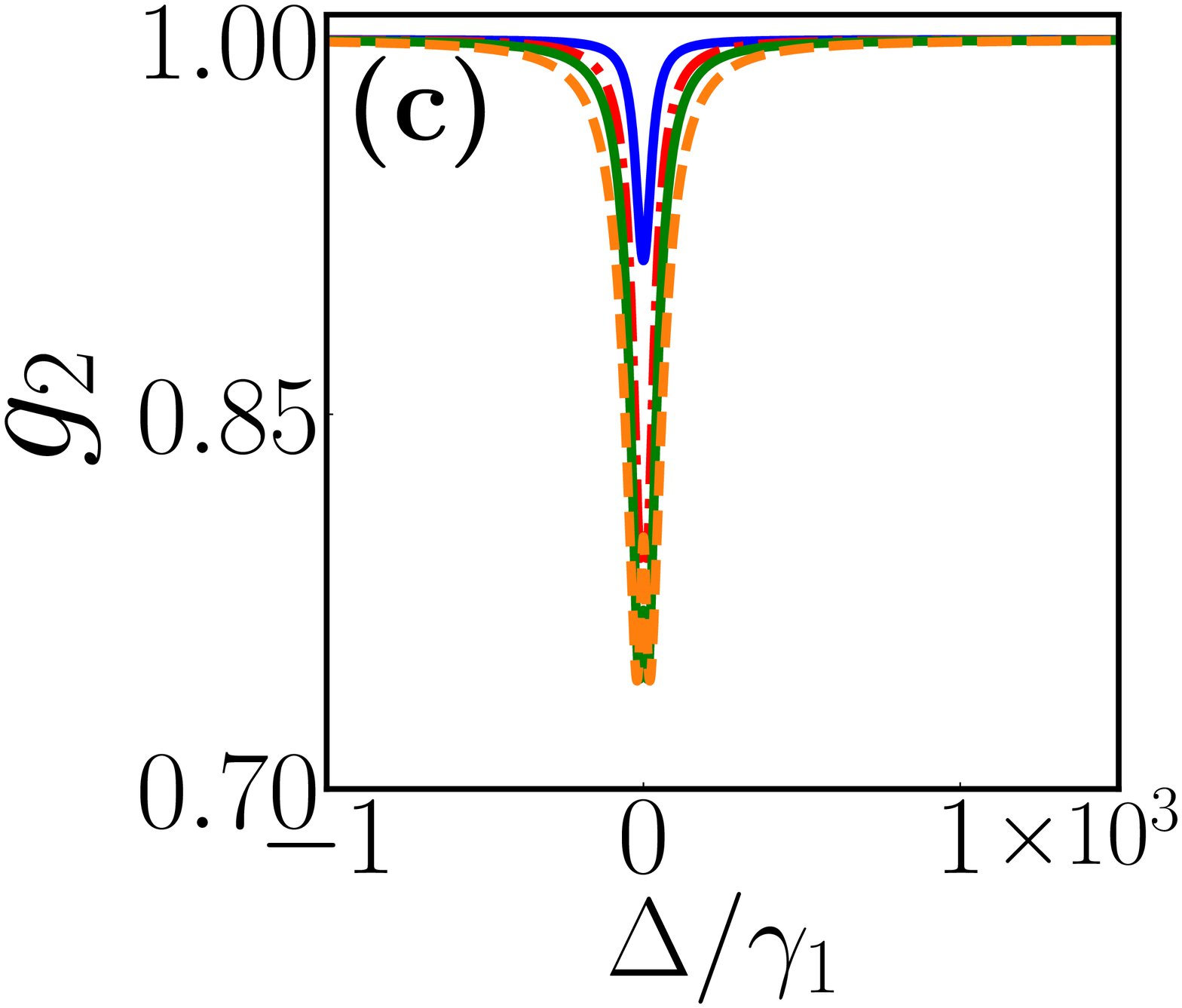}
 	\includegraphics[width=0.48\linewidth]{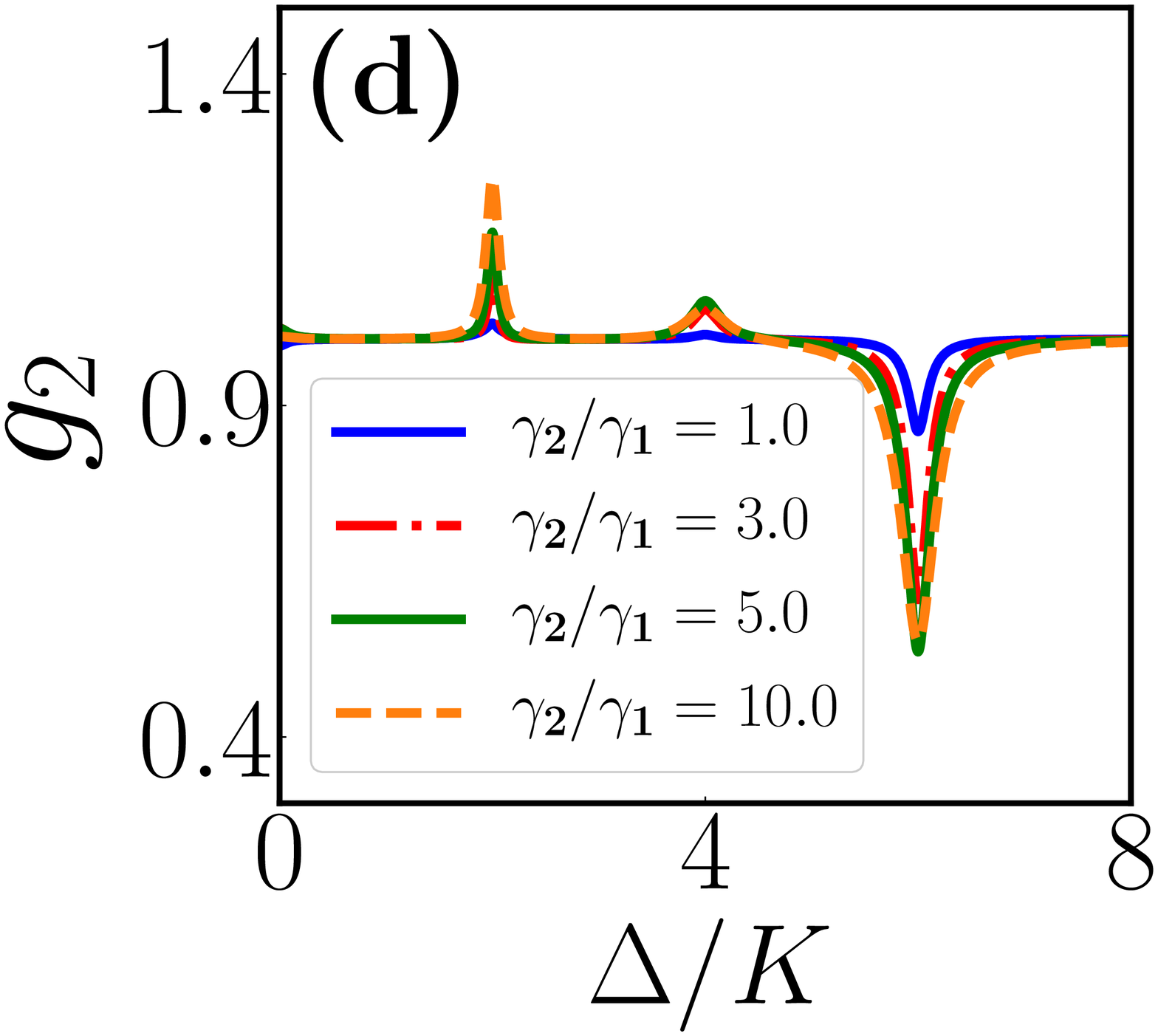}
 	\caption{ In Figs. \ref{corr2d}(a) and \ref{corr2d}(b) the correlation function $g_2$ shown against detuning frequency $\Delta$   for different coupling strengths (values of $\zeta$ given in the inset of Fig. \ref{corr2d}(b)) with damping rate $\gamma_2/\gamma_1=10$ and Figs. \ref{corr2d}(c) and \ref{corr2d}(d) for different damping rates (values of $\gamma_2/\gamma_1$ given in the inset of Fig. \ref{corr2d}(d)) with $\zeta=5.0$ for $K=0$ in Figs. \ref{corr2d}(a) and \ref{corr2d}(c) and $K=250\gamma_1$ in Figs. \ref{corr2d}(b) and \ref{corr2d}(d). }
 	\label{corr2d}
 \end{figure} 
In Figs. \ref{corr2d}(a) and \ref{corr2d}(b), we plot the second order correlation function $g_2$ as a function of coupling strength for $K=0$ and $K\neq0$ respectively.  In Figs. \ref{corr2d}(c) and \ref{corr2d}(d) we illustrate $g_2$ as a function of damping parameter $\gamma_2$, respectively for $K=0$ and $K\neq0$. For $K=0$, we observe that the phonons are anticorrelated and as such the second-order correlation function $g_2$ turns out to be less than $1$ for all values of $\zeta$. Hence the system exhibits antibunching and as we increase the coupling strength, the antibunching also increases as observed from Fig. \ref{corr2d}(a). We also notice a split in the antibuching dip, which increases with increasing damping parameter $\gamma_2/\gamma_1$ as shown in Fig. \ref{corr2d}(c). For lower values of $\gamma_2/\gamma_1$ we do not observe any split due to resonant absorption during the transition from $|0\rangle$ to $|1\rangle$. Due to this antibunching phenomenon we observe anticorrelation and as a result, phonon blockade in the synchronization peaks due to different phonon transitions between the Fock states in the coupling basis as shown in Fig. \ref{Fig1}. For $K\neq0$ we observe simultaneous bunching and antibunching at different resonances as illustrated in Figs. \ref{corr2d}(b) and \ref{corr2d}(d). At $\Delta=2K$ and $4K$ we observe bunching since $g_2>1$ which increases with increasing $\zeta$ (Fig. \ref{corr2d}(b)) and at $\Delta=6K$ we can see that $g_2<1$ and the phonons are antibunched. Because of the presence of this nonclassical effect in the system (\ref{mas_eq}), we observe a synchronization blockade in the quantum regime.
\section{Power Spectrum}\label{power_sp}
\par
\begin{figure}[!ht]
	\includegraphics[width=1.0\linewidth]{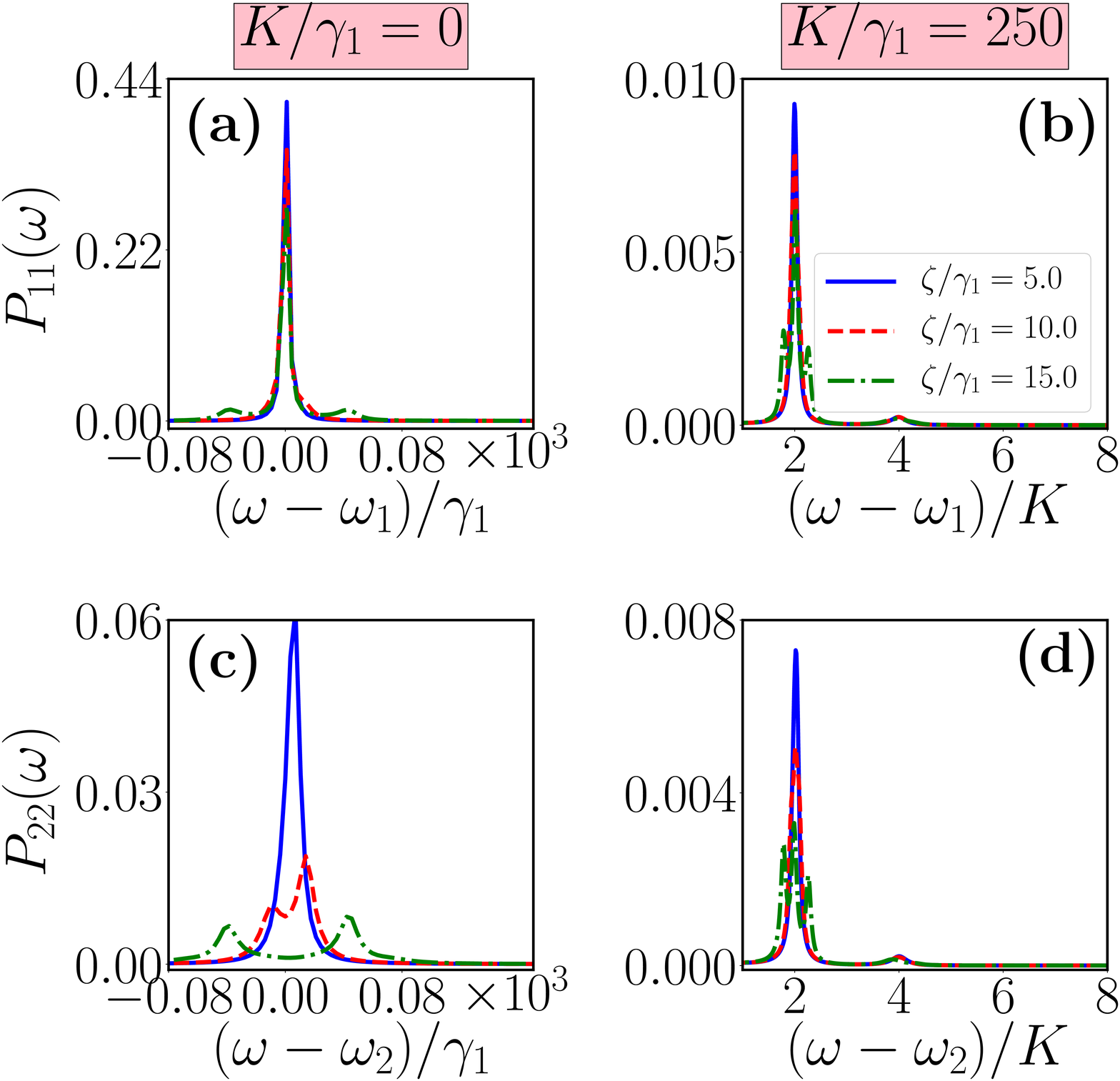}
	\caption{Power Spectra of first ($P_{11}(\omega)$) and second ($P_{22}(\omega)$) oscillators for different coupling strengths $\zeta$ (values given in the inset of Fig. \ref{sp_ze}(b)) for $\gamma_2/\gamma_1=10$ and $\Delta=0$.}
	\label{sp_ze}
\end{figure}

\begin{figure}[!ht]
	\includegraphics[width=1.01\linewidth]{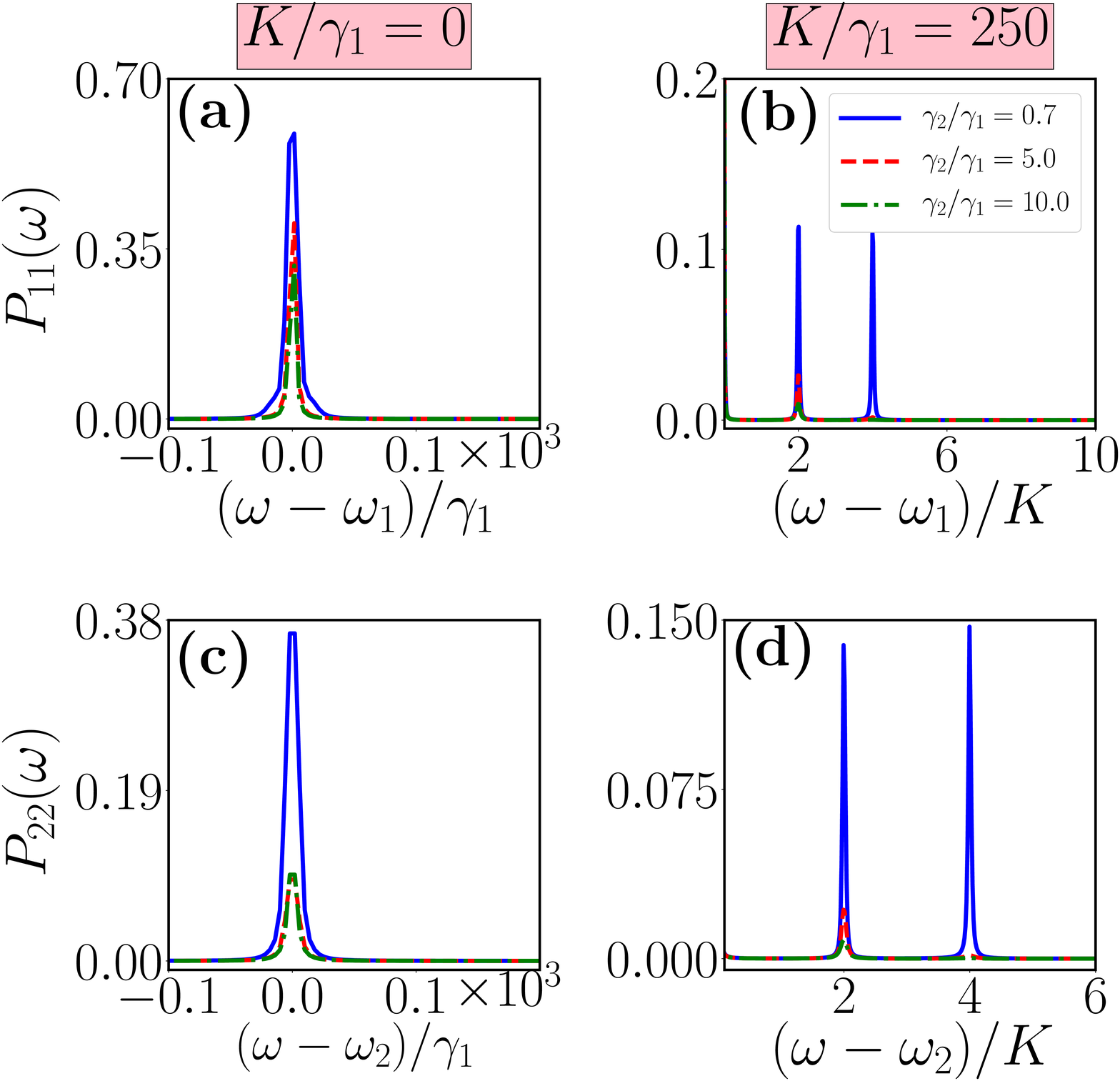}
	\caption{ Power Spectra of first ($P_{11}(\omega)$) and second ($P_{22}(\omega)$) oscillators for damping parameters $\gamma_2/\gamma_1$ (values given in the inset of Fig. \ref{sp_ze}(b)) for $\zeta=3.5\gamma_1$ and $\Delta=0$.. }
	\label{sp_da}
\end{figure}

 In the previous section we observed phonon-blockade in the phase-locking regime and as a consequence of the phonon-blockade, we observed antibunching effects in the same parametric region in the coupled quantum van der Pol oscillators. The phonon blockade occurs due to the appreciable excitation dependent frequency detuning present in the system. In the case of anharmonic quantum van der Pol oscillators, we observed multiple resonance synchronization peaks and simultaneous bunching and antibunching effects at different resonances as a result of anharmonic interaction. We can also investigate these attributes using the power mechanical spectrum defined by \cite{walter14}
\begin{equation}
P_{ii}(\omega)=\int_{-\infty}^{\infty}dt e^{i\omega t}\langle a_i^{\dagger}(t)a_i(0)\rangle,\qquad i=1,2. \label{power}
\end{equation}
which characterizes the frequency entrainment present in the system. Equation (\ref{power}) describes the energy spectrum of the oscillators. In Fig. \ref{sp_ze} we plot the power spectra $P_{11}(\omega)$ and $P_{22}(\omega)$ against the dimensionless frequency $\tilde{\omega_i}=(\omega-\omega_i)/\gamma_1$ ($i=1,2$) of first and second oscillator for different coupling strengths. In Figs. \ref{sp_ze}(a) and \ref{sp_ze}(c), for $K=0$, we observe spectral peaks at $\tilde{\omega_i}=0$ with $\omega_2=2\omega_1$. As we increase the coupling strength, the heights of the spectral peaks decrease and split to form a Mollow triplet in the case of first oscillator as demonstrated in Fig. \ref{sp_ze}(a). We observe a normal-mode split in the spectra ($P_{22}(\omega)$) of second oscillator when the coupling is strong as shown in Fig. \ref{sp_ze}. In the case of anharmonic oscillator we observe multiple spectral peaks corresponding to the resonance condition given in Eq. (\ref{res_con}) for increasing frequency ($\tilde{\omega_i}>0$)  for first and second oscillator as presented respectively in Figs. \ref{sp_ze}(b) and \ref{sp_ze}(d). Upon increasing the coupling strength we observe that the Mollow-triplet is formed in the spectral peaks of both oscillators at $\tilde{\omega}=0$ and $\tilde{\omega_i}=2$ as shown in Figs. \ref{sp_ze}(b) and \ref{sp_ze}(d). 
The effect of damping parameters in the spectrum of the oscillators is captured in Fig. \ref{sp_da}. By fixing the coupling strength at $\zeta=3.5\gamma_1$, we illustrate the spectral characteristics for increasing damping parameters for $K=0$ and $\Delta=0$ in Figs. \ref{sp_da}(a) and \ref{sp_da}(c). For $\gamma_2/\gamma_1\ll\zeta$, we observe that the Mollow triplet is formed in the spectra of first oscillator as shown in Fig. \ref{sp_da}(a). In the case of second oscillator, we observe a spectral peak at $\tilde{\omega_2}=0$. For $\gamma_2/\gamma_1>\zeta$, we observe a slight depression in the spectral peak of the second oscillator at $\tilde{\omega_2}=0$ as presented in Fig. \ref{sp_da}(c). For $\gamma_2/\gamma_1\gg\zeta$ the height of the spectral peaks get reduced and we do not observe any strong coupling characteristics. The spectral peaks of anharmonic oscillators for $K\neq0$ are illustrated in Figs. \ref{sp_da}(b) and \ref{sp_da}(d). For very low values of damping parameter we observe prominent spectral peaks of first oscillator for different resonance conditions (\ref{res_con}) at $\tilde{\omega_1}=0$, $\tilde{\omega_1}=2$ and $\tilde{\omega_1}=4$ as depicted in Fig. \ref{sp_da}(b). For second oscillator the spectral peaks are dominent for $\tilde{\omega_1}=0$, $\tilde{\omega_1}=2$ and less prominent for $\tilde{\omega_1}=4$. With increasing damping parameters the height of the spectral peaks of the anharmonic oscillators get suppressed as illustrated in Figs. \ref{sp_da}(b) and \ref{sp_da}(d).
\par As discussed previously the transition from ground $|0\rangle$ state and first excited state $|1\rangle$ is enhanced at resonance $\Delta=0$ for lower values of $\zeta$ and there we obtain single spectral peaks for the first and second oscillators for $K=0$ and multiple spectral peaks for $\tilde{\omega_i}>0$ in the anharmonic case ($K\neq0$). As we increase the coupling strength, the transitions between second and third excited states are far from resonance. Phonon-blockade occur there and as a result, we observe antibunching effects. The normal mode splitting in the form of the Mollow-triplet and two mode splitting in the spectral peaks of first and second oscillators respectively are the consequences of the aforementioned effects. In the strong coupling regime, formation of the Mollow-triplet is because of four different allowed transitions ($|3_{+}\rangle \to |2_{-}\rangle $, $|3_{+}\rangle \to |2_{+}\rangle $, $|3_{-}\rangle \to |2_{-}\rangle $ and $|3_{-}\rangle \to |2_{+}\rangle $) between the second and third excited states and the sideband frequencies occur at $\tilde{\omega_i}\pm\omega_{j\pm}$ where $\omega_{j\pm}$ ($j=1,2$) are the frequencies of nondegenerate states $|2_{\pm}\rangle$ and $|3_{\pm}\rangle$ respectively. The two peak normal mode splitting is due to blocked transition of second and third excited states ($|2_{\pm}\rangle$ and $|3_{\pm}\rangle$) with peaks occuring at $\tilde{\omega_i}\pm\omega_{j\pm}$. The allowed transition between the Fock states depends on the  two phonon loss rate ($\gamma_2/\gamma_1$) of the quantum van der Pol oscillators. 

\subsection{Experimental Realization}
\par The system represented by Eq. (\ref{mas_eq}) can be experimentally achieved via trapped ion setup by implementing side-band transitions for two motional modes of frequencies $\omega_0$ and $2\omega_0$ respectively \cite{lee}. Then driving side-band transition of both the modes off-resonantly from an excited state we can obtain the coupling described by the Hamiltonian in Eq. (\ref{int_h}). The system can be characterized by Wigner-parity function. The interaction brings out a strong coupling between the modes which is a desirable property in quantum information processing \cite{ding17}. Large Kerr nonlinearities can be engineered in trapped ions \cite{bruder, grimm, home}. The nonlinearly coupled quantum van der Pol oscillator can also be realized in a cavity optomechanical system \cite{walter}. The nonlinear interaction between the mechanical van der Pol oscillators can be realized by quadratically coupling the ``membrane-in-the-middle" setup \cite{walter, xie}. The cavity mode $c$ can be added to nonlinearly coupled mechanical oscillator and driven with laser at  frequency $\omega_p$. The total Hamiltonian is given by
\begin{eqnarray}
H=\omega_0c^{\dagger}c+E e^{-i\omega_pt}c^{\dagger}+E^*e^{i\omega_pt}c\nonumber\\+\sum_{i=1,2}\omega_ia_i^{\dagger}a_{i}+g_ic^{\dagger}c(a_i+a_i^{\dagger})^2,
\end{eqnarray}
where $\omega_0$ is the cavity frequency, $g_i$ is the optomechanical coupling strength and $E$ is the driving strength of the laser. Large Kerr anharmonicities are difficult to realize in optomechanical setup but it can be realized in hybrid systems \cite{amitai2, rips, rimberg}.
\section{Conclusion}\label{conc}
In this work we have investigated the synchronization dynamics of nonlinearly coupled quantum van der Pol oscillators and those of anharmonic self-oscillators. We have shown that the system exhibits certain novel features in the quantum domain which are not present in the classical domain. We have identified that due to anharmonicity of the nonlinear coupling, the system exhibits synchronization blockade in the phase-locking regime. We have also shown a quantized phase-locking behaviour in nonlinearly coupled anharmonic self-oscillators which comes out due to the heterogeneity that is present in the system. We have illustrated that the phonon blockade in the system, which arises due to anticorrelation between the oscillators, increases with coupling strength. Further, we have also demonstrated that due to negative correlation between the oscillators, the system shows antibunching effects in the phase-locking parametric regime. In the case of anharmonic self-oscillators, we have observed simultaneous correlation and anticorrelation between the oscillators at different resonance peaks which also leads to simultaneous bunching and antibunching in the system. We have identified that the limit cycle of the anticorrelated oscillator shrinks with increasing coupling strength in the Wigner distribution function. We have also shown these attributes in the frequency entrainment of the system. The system shows the normal-mode splitting for higher values of coupling strength, a feauture of strong nonlinear interactions. We have also observed the Mollow triplet due to multi-phonon transitions in the system with increasing coupling strengths. 
\par Phonon and Photon blockade in the quantum systems has been an important topic of research since it is a pure quantum effect that leads to antibunching effects in the system. In single photon detectors, photon correlation is an important tool and has applications in quantum information processing such as quantum teleportation \cite{boschi, riebe, muller}, quantum cryptography \cite{ekert, souja} and so on. Several theoretical and experimental studies have been conducted for the detection of phonon blockade in optomechanical systems \cite{xu,lee3,zhou,gerace,xie}, nanomechanical resonators \cite{miran}, optical cavity with one trapped atom \cite{birna}, cavity QED \cite{liao}, and superconducting microwave resonator \cite{vaneph}. Quantum van der Pol oscillator also provides a feasible phonon source and our studies can help in the realization of phonon detection and quantum information tasks.
\section*{Acknowledgements}
NT thanks National Board for Higher Mathematics, Government of India, for providing the Senior Research Fellowship under the Grant No. 02011/20/2018 NBHM (R.P)/R\&D II/15064 to carry out this work. The work of MS forms part of a research project sponsored by Science and Engineering Research Board,  Government of India, under the Grant No. CRG/2021/002428.

\end{document}